\documentstyle[aps,rotate,multicol,epsf]{revtex}
\def\beginwide{
        \end{multicols} \vspace*{-0.5cm} \noindent
        \rule{3.5in}{.1mm}\rule{.1mm}{5mm} \widetext \medskip }
\def\beginwidetop{
        \end{multicols} \vspace*{-0.5cm} \noindent
        \widetext \medskip }
\def\endwide{
        \hspace*{3.35in}~\rule[-5mm]{.1mm}{5mm}\rule{3.5in}{.1mm}
        \begin{multicols}{2} \vspace*{-1.0cm} \noindent }
\def\endwidebottom{
        \begin{multicols}{2} \vspace*{-1.0cm} \noindent }

\newcommand{\beq}{\begin{equation}}
\newcommand{\eeq}{\end{equation}}
\newcommand{\bea}{\begin{eqnarray}}
\newcommand{\eea}{\end{eqnarray}}

\draft

\begin{document}

\title{ An Integrable Version of Burgers Equation in 
``Magnetohydrodynamics''}
\author{P. Olesen\footnote{email: polesen@nbi.dk} }
\address{Niels Bohr Institute,
Blegdamsvej 17, DK-2100 Copenhagen {\O}, Denmark }
\date{\today}
\maketitle

\begin{abstract}
It is pointed out that for the case of (compressible) 
magnetohydrodynamics (MHD) 
with the fields $v_y(y,t)$ and $B_x(y,t)$ one can have equations of
the Burgers type which are integrable. We discuss the solutions. It turns
out that the propagation of the non-linear effects is governed by the initial
velocity (as in Burgers case) as well as by the initial Alfv\'en velocity.
Many results previously obtained
for the Burgers equation can be transferred to the MHD case. We also
discuss equipartition $v_y=\pm B_x$. It is shown that an initial localized 
small scale magnetic field will end up in fields moving to the left and the 
right, thus transporting energy from smaller to larger distances.
\end{abstract}

\begin{multicols}{2}

\vskip0.3cm
The Burgers equation \cite{burgers} has been much studied. It can be applied 
to a variety of
phenomena, see e.g. \cite{whitham},\cite{gurbatov}, \cite{zeldovich}. 
Although this equation 
satisfies a number of properties which are similar to hydrodynamics, it
is known to be integrable. Hence the Burgers equation does not have the
properties characterizing chaotic dynamical systems. However, to some
extent such properties may be simulated by random boundary conditions 
\cite{sinai},\cite{she}. Also, the long time behavior of decaying solutions of 
the Burgers equation with an initial velocity which is  homogeneous
and Gaussian has been studied and many interesting properties of the
spectrum have been found \cite{frisch}. Also, a general statistical
theory of the Stochastic Burgers equation in the inviscid limit has
been developed \cite{weinan}. For a recent review of ``Burgulence''
we refer to the paper by Frisch and Bec \cite{bec}. 
These results make it clear that a number
of properties of the Burgers equation are highly non-trivial. With this in mind
we present an integrable generalization of Burgers equation. The
new equations are related to magnetohydrodynamics (MHD) in
a way which is analogous to the relation between the Burgers equation and 
the Navier-Stokes equations. Essentially all properties found for the Burgers
equation can be applied to the new equations, but they also contain
some new features.

MHD in 1+1 dimension has been considered by Thomas \cite{thomas} long time 
ago. The fields are the velocity $v_x(x,t)$ and the magnetic field $B_x(x,t)$,
which satisfy
\begin{eqnarray}
&&\partial_t v_x+v_x~\partial_x v_x =B_x~\partial_x B_x+\nu~\partial^2_x v_x,
\nonumber \\
&&\partial_t B_x+v_x~\partial_x B_x =B_x~\partial_xv_x+(1/\sigma)~\partial^2_x 
B_x.
\label{thomas}
\end{eqnarray}
The first of these equations is similar to the Burgers equation.
Both equations are modeled from the incompressible MHD equations.
However, since incompressibility leads to triviality in 1+1 dimensions,
the equation div ${\bf v}=0$ is not enforced, and the total (including the
magnetic) pressure is taken to vanish. Also,
in these equations it is implicit that variation of the density is
disregarded. Similarly, the equation div ${\bf B}=0$ is not satisfied.
It has been shown by Passot \cite{passot} that the above equations are not 
integrable. For a review of some of the consequences of these equations, we 
refer to ref. \cite{leshouches}.

In the following we consider a different form of
1+1 dimensional MHD, where there is really two dimensions,
$x$ and $y$, but where the fields are restricted to depend only on $y$. We 
restrict ourselves to the fields $B_x(y,t)$ and $v_y(y,t)$. The
equation of motion\footnote{In eqs. (\ref{v}) and (\ref{B}) the density $\rho$
and the vacuum permeability $\mu_0$ should occur. Thus, ${\bf (\nabla\times
B)\times B}$ on the right hand side should be multiplied by $1/\mu_0\rho$.
We assume a constant
$\rho$. Then the following rescalings ``remove'' $\rho$ and $\mu_0$: 
$\sigma\mu_0\rightarrow\sigma$ and ${\bf B}\rightarrow {\bf B}/
\sqrt{\mu_0 \rho}$.}
\begin{equation}
\partial_t{\bf v}+({\bf v\nabla}){\bf v}=({\bf \nabla\times B})\times{\bf B}
+\nu~\nabla^2{\bf v}
\end{equation}
then becomes
\begin{equation}
\partial_t v_y+v_y\partial_yv_y=-B_x\partial_yB_x+\nu~\partial_y^2v_y.
\label{v}
\end{equation}
Similarly, the equation
\begin{equation}
\partial_t {\bf B}={\bf \nabla\times (v\times B)}+(1/\sigma)~ \nabla^2 {\bf B},
\end{equation}
where $\sigma$ is the conductivity of the fluid, becomes
\begin{equation}
\partial_t B_x=-\partial_y (v_yB_x)+(1/\sigma)~\partial_y^2B_x.
\label{B}
\end{equation}
The coupled set of equations (\ref{v}) and (\ref{B}) can be interpreted
by saying that (\ref{v}) is a Burgers equation with a magnetically
generated pressure, governed by (\ref{B}). Like in the case of Thomas's
1+1 dimensional equations, div ${\bf v}=0$ is not satisfied,
and we have also disregarded a possible variation of the
density. Notice however that div$~{\bf B}=0$ is satisfied in our approach
and that the magnetic pressure $B_x^2/2$ is kept.

The main difference between our equations and those discussed by Thomas is that
his equation (\ref{thomas}) includes a term $B_x\partial_x v_x$ which stretches
the magnetic field lines, and which competes with the term $v_x\partial_xB_x$ 
which (in higher dimensions) breaks or twists the field lines. On the other
hand, in our case the magnetic field is divergence free, and we included the 
magnetic pressure.

The reduced model (\ref{thomas}) introduced by Thomas is in correspondence
to special subclasses of of solutions to the full MHD equations. In our case
the equations (\ref{v}) and (\ref{B}) are of course derived from the
real MHD equations, but they do not directly correspond to known subclasses
of solutions of the MHD equations. However, the reduced model proposed
in this paper may have some physical interest, which we shall discuss at the 
end of this paper after having given the analytic solution of eqs.
(\ref{v}) and (\ref{B}). It turns out that the model predicts that
magnetic energy which is initially localized at small scales, is moved to 
large distances by the non-linear dynamics.

Conservation of energy can easily be checked from eqs. (\ref{v}) and
(\ref{B}). With
\begin{equation}
E_{\rm tot}(t)=\int_{-\infty}^\infty~dy ~\left(v_y^2+B_x^2\right),
\end{equation}
one has
\begin{eqnarray}
&&\frac{dE_{\rm tot}(t)}{dt}=2 \int_{-\infty}^\infty~dy\nonumber \\
&&\times\left(-\frac{1}{3}
\partial_yv_y^3-\partial_y(v_yB_x^2)+\nu~ v_y\partial_y^2v_y+\frac{1}{\sigma}~
B_x\partial_y^2B_x\right).
\end{eqnarray}
Assuming no ``diffusion at infinity'', i.e. assuming that $v_y$ and
$B_x$ vanish for $y\rightarrow\pm\infty$, then
\begin{equation}
 \frac{dE_{\rm tot}(t)}{dt}=-2 \int_{-\infty}^\infty~dy~\left(\nu~
(\partial_yv_y)^2+\frac{1}{\sigma}~(\partial_yB_x)^2\right).
\end{equation}
Thus energy is conserved in the limit $\nu,1/\sigma\rightarrow 0$.

The idea is now to compare the equations (\ref{v}) and (\ref{B}) to the
well known solution of Burgers equation found by Hopf \cite{hopf} and 
Cole \cite{cole}, where the diffusive terms in these
equations are included. We can show that if
\begin{equation}
\nu =1/\sigma,
\label{special}
\end{equation}
then the equations are integrable. We do not know if the equations are still
integrable if $\nu\neq 1/\sigma$, and/or if variations of the density $\rho$ 
are included according to the conservation equation
\begin{equation}
\partial_t\rho +\partial_y(\rho v_y)=0.
\end{equation}
Of course, the full set of equations can be studied numerically.

In the following we shall consider the case where (\ref{special}) is
satisfied. It should be emphasized that
this assumption is not supposed to represent a realistic estimate of
the magnetic Prandtl number $P_m=\sigma\nu$, which e.g. for liquid metals is
of the order\footnote{It can be mentioned that in numerical 
simulations of the Earth's dynamo with Prandtl number $\sim 10^{-6}$
one actually uses a value of $P_m\sim 0.1$ (not so far from one!), since this 
is what is numerically tractable. See e.g. \cite{mhd}.} $10^{-5}$. Our excuse 
for having a 
Prandtl number equal to one is primarily that this allows a non-trivial 
solution of eqs. (\ref{v}) and (\ref{B}). Furthermore, one space dimension 
is anyhow not realistic. 

Adding and subtracting
eqs. (\ref{v}) and (\ref{B}) we obtain in the special case (\ref{special})
\begin{equation}
\partial_t (v_y+B_x)+\frac{1}{2}\partial_y(v_y+B_x)^2=\nu ~\partial_y^2
(v_y+B_x),
\label{v+B}
\end{equation}
and
\begin{equation}
\partial_t (v_y-B_x)+\frac{1}{2}\partial_y(v_y-B_x)^2=\nu ~\partial_y^2
(v_y-B_x).
\label{v-B}
\end{equation}
These two equations are of the Burgers type. 

We remind the reader that the solution of Burgers equation
\begin{equation}
\partial_t u+u\partial_y u=\nu \partial_y^2 u
\end{equation}
is
\begin{equation}
u(y,t)=\frac{1}{t}(y-\bar{a}(y,t)), 
\end{equation}
where
\begin{eqnarray}
&&\bar{a}(y,t)=\nonumber \\
&&\int_{-\infty}^\infty~ae^{-\frac{(y-a)^2}{4\nu t}+\frac{1}{2\nu}
\psi (a)}da\left/\int_{-\infty}^\infty~e^{-\frac{(y-a)^2}{4\nu t}+
\frac{1}{2\nu}\psi(a)}da. \right.
\label{burgers}
\end{eqnarray}
Here $u(y,t=0)=-\partial_y\psi (y)$. For the case where $\nu =0$ we have
\begin{equation}
u(y,t)=u(b(y,t),t=0),
\label{characteristic}
\end{equation}
where $b(y,t)$ solves the implicit equation
\begin{equation}
b(y,t)=y-t~u(b(y,t),0).
\label{bb}
\end{equation}
Here $b(y,t)$ can be interpreted as a Lagrangian coordinate, with $b(y,0)=y$
for $t=0$. 
This solution can be obtained by the methods of characteristics or from the
saddle point in eq. (\ref{burgers})  for $\nu\rightarrow 0$. In this case
$b(y,t)\rightarrow \bar{a}(y,t)$. In the case where there are more
than one solution of eq. (\ref{bb}) for $b(y,t)$ one should take the
solution which maximizes the expression
\begin{equation}
-\frac{(y-b)^2}{2t}+\psi (b)
\end{equation}
with respect to $b$,
as is obvious from the saddle point expansion of (\ref{burgers}).

For the MHD case the solution can be found in terms of the initial values
\begin{eqnarray}
&&v_y(y,t=0)\equiv -\partial_y\psi_{v0}(y),\nonumber \\
&&B_x(y,t=0)\equiv -\partial_y\psi_{B0}(y).
\end{eqnarray}
The Hopf-Cole solution of Burgers' equation then gives
\begin{eqnarray}
v_y(y,t)&=&\frac{1}{t}\left[y-\frac{1}{2}\left(\bar{a}_{v+B}(y,t)+\bar{a}_{v-B}
(y,t)\right)\right]\nonumber \\
&=&-\frac{1}{2}(\bar{\psi'}_{v0+B0}(y,t)+\bar{\psi'}_{v0-B0}
(y,t)),
\label{solvev}
\end{eqnarray}
and
\begin{eqnarray}
B_x(y,t)&=&-\frac{1}{2t}\left(\bar{a}_{v+B}(y,t)-\bar{a}_{v-B}
(y,t)\right)\nonumber \\
&=&-\frac{1}{2}(\bar{\psi'}_{v0+B0}(y,t)-\bar{\psi'}_{v0-B0}
(y,t)).
\label{solveB}
\end{eqnarray}
Here
\begin{eqnarray}
&&\bar{a}_{v\pm B}(y,t)=\nonumber \\
&&\int_{-\infty}^{\infty}ae^{-\frac{(y-a)^2}{4\nu t}+
\frac{1}{2\nu }(\psi_{v0}(a)\pm\psi_{B0}(a))}da\nonumber \\
&&\left[\int_{-\infty}^{\infty}
e^{-\frac{(y-a)^2}{4\nu t}+\frac{1}{2\nu }(\psi_{v0}(a)
\pm\psi_{B0}(a))}da\right]^{-1},
\label{solution}
\end{eqnarray}
and similarly
\begin{eqnarray}
&&\bar{\psi'}_{v0\pm B0}(y,t)=\nonumber \\
&&\int_{-\infty}^{\infty}(\psi_{v0}'(a)\pm\psi_{B0}'(a))e^{-\frac{(y-a)^2}
{4\nu t}+
\frac{1}{2\nu }(\psi_{v0}(a)\pm\psi_{B0}(a))}da\nonumber \\
&&\left [\int_{-\infty}^{\infty}
e^{-\frac{(y-a)^2}{4\nu t}+\frac{1}{2\nu }(\psi_{v0}(a)\pm\psi_{B0}(a))}da
\right]^{-1}.
\label{solution2}
\end{eqnarray}
The last forms of eqs. (\ref{solvev}) and (\ref{solveB}) follow from 
\begin{equation}
\int_{-\infty}^{\infty}\frac{\partial}{\partial a}~e^{-\frac{(y-a)^2}
{4\nu t}+
\frac{1}{2\nu }(\psi_{v0}(a)\pm\psi_{B0}(a))}~da=0.
\end{equation}
In case where $\nu$ is taken to nearly vanish, the resulting saddle point
simplifies the solutions (\ref{solvev}) and (\ref{solveB}), and the $\bar{a}$'s
are replaced by solutions of the equations
\begin{equation}
\bar{a}_{v\pm B}=y+t\left(\psi'_{v0}(\bar{a}_{v\pm B})\pm\psi'_{B0}
(\bar{a}_{v\pm B})\right).
\end{equation}
These solutions can also be obtained by solving the original equations without 
diffusion ($\nu=1/\sigma =0$) by the methods of characteristics. These 
solutions can be written in a form analogous to eq. (\ref{characteristic}),
\begin{eqnarray}
&&B_x (y,t)=\frac{1}{2}\left( B_x(b_+(y,t),0)+ B_x(b_-(y,t),0)\right.
\nonumber \\
&&\left.+v_y(b_+(y,t),0)-v_y(b_-(y,t),0)\right),
\label{B2}
\end{eqnarray}
and
\begin{eqnarray}
&&v_y(y,t)=\frac{1}{2}\left(v_y(b_+(y,t),0)+v_y(b_-(y,t),0)\right.\nonumber \\
&&\left.+B_x(b_+(y,t),0)- B_x(b_-(y,t),0)\right),
\label{v2}
\end{eqnarray}
where $b_{\pm}  (y,t)$ solve the equations
\begin{equation}
b_{\pm} (y,t)=y-t~v_y(b_{\pm}(y,t),0)\mp t ~B_x(b_{\pm}(y,t),0).
\label{b}
\end{equation}
Like in Burgers case $b_\pm (y,0)=y$. Also, there is
a simple Lagrangian interpretation of eq. (\ref{b}), since the right hand
side involves the initial velocity subtracted or added to the usual 
Alfv\'en velocity $B_x(b_\pm,0)$, where it should be
remembered that the constant $1/\sqrt{\mu_0\rho}$ was absorbed in $B_x$.
In the saddle point of eq. (\ref{solution}) for $\nu\rightarrow 0$ we again 
have $\bar{a}_{v\pm B}(y,t)\rightarrow b_{\pm}(y,t)$, with $b_\pm$ given by the
dominant saddle point, as discussed below eq. (\ref{bb}). It is of
course easy  to show directly that eqs. (\ref{B2}) and (\ref{v2}) 
satisfy the original equations (\ref{v}) and (\ref{B}) with $\nu =1/\sigma=0$.

From the solutions (\ref{B2}) and (\ref{v2}) one can read off a few simple 
properties. If the initial magnetic field $B_x(y,t=0)$ vanishes no magnetic
field is generated at other times. This is already quite obvious from the 
original equations
(\ref{v}) and (\ref{B}). To see this result from the solution by 
characteristics we notice that if $B_x$ vanishes at $t=0$ it follows from
(\ref{b}) that $b_+=b_-$, and eq. (\ref{B2}) then gives $B_x(y,t)=0$. The
velocity field will then behave as a solution of the ``pure'' Burgers equation
for $v_y$.

A less trivial case is when the initial velocity field vanishes,
\begin{equation}
v_y(y,t=0)=0.
\end{equation}
Then we obtain both a magnetic and a velocity field as a consequence of
the dynamics, namely
\begin{eqnarray}
B_x (y,t)&=&\frac{1}{2}\left( B_x(b_+(y,t),0)+ B_x(b_-(y,t),0)\right),
\nonumber \\
v_y(y,t)&=&\frac{1}{2}\left(B_x(b_+(y,t),0)- B_x(b_-(y,t),0)\right).
\label{1}
\end{eqnarray}
with
\begin{equation}
b_{\pm} (y,t)=y\mp t ~B_x(b_{\pm}(y,t),0).
\label{2}
\end{equation}
We see that if the initial magnetic field is constant, no velocity field is 
generated. However, in general a varying initial magnetic field is able to 
generate a velocity field. 

In MHD it has often been discussed whether there is equipartition, 
$v_y=\pm B_x$, after a long time. In realistic MHD simulations one
does not in all cases find equipartition. A recent study \cite{axel} finds
that in nonhelical hydromagnetic turbulence in the inertial range the magnetic 
energy exceeds the kinetic energy by a factor of 2 to 3. The helical case 
has recently been discussed in reference \cite{axel2}. 
In our case it follows rather trivially that if
the initial fields satisfy equipartition, then this will be true for
all times. In
general, we see from eqs. (\ref{B2}) and (\ref{v2}) that equipartition 
in the exact sense requires
\begin{eqnarray}
&&B_x(b_-(y,t),0)=v_y(b_-(y,t),0)~~{\rm and}\nonumber \\
&& B_x(b_+(y,t),0),v_y(b_+(y,t),0)~~{\rm arbitrary},\nonumber \\
&&{\rm or}\nonumber \\
&&B_x(b_+(y,t),0)=v_y(b_+(y,t),0)~~{\rm and}\nonumber \\
&&B_x(b_-(y,t),0),v_y(b_-(y,t),0)~~{\rm arbitrary},
\label{sveske}
\end{eqnarray}
for $v_y=B_x$ and $v_y=-B_x$, respectively. In the first case it follows
from eq. (\ref{b}) that $b_-=y$. Hence from the first line in (\ref{sveske})
it follows that $B_x(y,0)=v_y(y,0)$, so the initial fields are equal. 
Thus, equipartition requires very special initial fields and is not
possible for general initial conditions. 

The considerations above do not, however, answer the question concerning
equipartition after some time has passed. This requires the study of (\ref{2})
for $b_\pm$ after the passage of
a sufficient time. Let us consider the case $v_y(y,0)=0$ and let us
take the initial field $B_x(y,0)$ to be localized in in a domain $D$ in $y$. 
Then eq.(\ref{2}) shows that $b_\pm$ receive non-trivial contributions from
$b_\pm\in D$, and these contributions have different signs, i.e. the
non-trivial part of $y-$space move to
the right and to the left, so the original domain $D$ splits up into
right and left moving domains $y\in D_R$ and $y\in D_L$. After some 
time has passed these domains have no overlap. Using the result (\ref{1}) 
we then see that $B_x(y,t)$ also moves to the left (right) with value
$B_x(b_-,0)/2$ ($B_x(b_-,0)/2$). Thus the value of the magnetic field has
decreased by a factor of 2. At the same time it follows from (\ref{1}) 
that the velocity has increased from zero to $\pm B_x(b_\pm)/2$. Therefore
one has equipartition.  

If the initial velocity is non-vanishing it is more difficult to estimate the
result from eqs. (\ref{1}) and (\ref{2}). We shall therefore ask what
is the tendency after a short time. We start by considering initial fields 
which are proportional,
\begin{equation}
B_x(y,0)=\lambda v_y(y,0),
\end{equation}
where $\lambda$ is some parameter. We now want to solve eqs. (\ref{B2}), 
(\ref{v2}), and (\ref{b}) perturbatively for small $t$. Assuming that $y$
is not too close to zero, eq. (\ref{b}) can be solved approximately,
\begin{equation}
b_\pm (y,t)\approx y-(1\pm\lambda)tv_y(y,0),
\end{equation}
where the assumption that $y$ does not vanish is needed in order that the 
second term on the right-hand side is small relative to $y$. The fields in
(\ref{B2}) and (\ref{v2}) can then be expanded, using
\begin{equation}
v_y(b_\pm,0)\approx v_y(y,0)\left(1-t(1\pm \lambda)~\partial_yv_y(y,0)\right),
\end{equation}
to obtain the results
\begin{eqnarray}
&&B_x(y,t)\approx\lambda v_y(y,0)(1-2t~\partial_yv_y(y,0)),\nonumber \\
&&v_y(y,t)\approx v_y(y,0)(1-t(1+\lambda^2)~\partial_yv_y(y,0)).
\label{lummer}
\end{eqnarray}
We see that the first order changes in the fields do not preserve the 
initial proportionality.
If we start with a magnetic field which is much smaller than
the velocity, $\lambda \ll 1$, then the relative change is larger
in $B_x$ than in $v_y$. If, on the other hand
$\lambda\sim 1$, i.e. the initial fields are comparable, then the corrections
found in (\ref{lummer}) are of the same order. If $\lambda$ is large, 
i.e. the initial magnetic field is much larger than the initial velocity,
then the correction to the velocity is larger than the correction to the 
magnetic field. Thus it seems that the non-linearity in the basic equations
tend to increase the smallest initial field, for small times. 

We mention that the contributions to the energy from the $O(t)$ corrections 
in (\ref{lummer}) is proportional to
\begin{equation}
t\int_{-\infty}^\infty dy~v_y(y,0)^2~\partial_yv_y(y,0)=\frac{t}{3}v_y(y,0)^3
\left|_{-\infty}^\infty, \right.
\end{equation}
which vanishes since $v_y(y,0)$ must approach zero at infinity in order
that the energy is finite. Thus the corrections in (\ref{lummer}) give
no contribution to the total energy. In this argument we have disregarded that 
the perturbation may not be valid very close to $y=0$.

We have also investigated the situation numerically, taking the initial fields
$v_y(y,0)=\sin y$ and  $B_x(y,0)=\lambda v_y(y,0)$, with $-\pi /2<y<\pi /2$. 
We find that after a long time the fields fluctuate considerably.
In general, there is no equipartition, except for $\lambda =1$. The 
(fluctuating) ratio
\begin{equation}
R=\frac{B_x^2}{v_y^2+B_x^2}
\end{equation}
is maximally of order 0.006 for $\lambda =0.2$ after a long time ($t=14$). For 
$\lambda =0.9$ one gets a maximum $R$ value around 0.37.
Finally, for $\lambda =1.1$ one gets a maximal $R$ around 0.5. Of course, 
if $\lambda =1$ the numerical calculations give equipartition with $R=0.5$
for all $y$. 

It is clear that the usual Burgers shock waves are present in our case too. 
From eq. (\ref{b}) one has that the derivatives of $b_\pm (y,t)$
become infinite for
\begin{equation}
t=t(b_\pm)=-\frac{1}{v'_y (b_\pm,0)\pm B'_x (b_\pm,0)}.
\end{equation}
In general there will actually be more shocks than in the ``pure'' Burgers 
case, since derivatives of the solutions (\ref{v2}) and (\ref{B2}) contain
$\partial_y b_+$ as well as $\partial_y b_-$ which are infinite at $t(b_+)$ as 
well as $t(b_-)$.

In our case the usual conservation form of Burgers equation are generalized to
\begin{equation}
\partial_tv_y=\partial_y\left(-\frac{1}{2}v_y^2-\frac{1}{2}B_x^2+\nu~
\partial_y v_y\right).
\end{equation}
Moreover, we also have
\begin{equation}
\partial_t B_x=\partial_y \left(-v_yB_x+\frac{1}{\sigma}~\partial_y B_x\right).
\end{equation}
In the last equation one needs of course to replace $1/\sigma$ by $\nu$ in 
order to apply the solutions found in this note.

As already mentioned, the many highly non-trivial properties of the
Burgers equation are shared by the solutions for $v_y$ and $B_x$. This
follows simply from the fact that eqs. (\ref{v+B}) and (\ref{v-B}) are
of the Burgers type, so any property previously derived can be
applied to the fields $v_y+B_x$ and $v_y-B_x$. Thus, except for accidental
cancelations these properties also hold for the fields $v_y$ and $B_x$.
To give an example, consider the correlation function
\begin{equation}
C(y,t)=<v_y(y,t)v_y(0,t)+B_x(y,t)B_x(0,t)>,
\end{equation}
where we take homogeneous random fields $v_y(y,t)$ and $B_x(y,t)$ with
ensemble averages.
$C(y,t)$ can be obtained from the sum of the correlation functions
\begin{equation}
<(v_y(y,t)+B_y(y,t))(v_y(0,t)+B_y(0,t))>
\label{kuk} 
\end{equation}
and
\begin{equation}
<(v_y(y,t)-B_y(y,t))(v_y(0,t)-B_y(0,t))>.
\label{kkuk}
\end{equation}
Each of these two correlation functions contain fields that are solutions of
the Burgers equations (\ref{v+B}) and (\ref{v-B}). Hence we can use known
results from Burgulence, for example from references \cite{frisch} and 
\cite{bec}, to obtain information on $C(y,t)$. The
total energy spectrum is then given by
\begin{equation}
E_{\rm tot}(k,t)=\frac{1}{2\pi}\int_{-\infty}^\infty C(y,t)~e^{iky}~dy.
\end{equation}
The total kinetic and magnetic energy is then given by
\begin{equation}
E_{\rm tot}(t)\equiv <v_y(y,t)^2+B_x(y,t)^2>=\int_{-\infty}^\infty 
E_{\rm tot}(k,t)~dk.
\end{equation}
By subtracting eqs. (\ref{kuk}) and (\ref{kkuk}) we can also obtain 
the correlator
\begin{eqnarray}
&&D(y,t)=<B_x(y,t)v_y(0,t)+B_x(0,t)v_y(y,t)>\nonumber \\
&&=\int_{-\infty}^\infty F(k,t)~e^{-iky}dk.
\end{eqnarray}
The function $F$ is related to the Lorentz force
\begin{equation}
F(t)\equiv 2<v_y(y,t)B_x(y,t)>=\int_{-\infty}^\infty F(k,t)~dk.
\end{equation}
It is now possible to repeat for example the arguments in \cite{frisch} to
obtain information on $C(y,t)$ and $F(y,t)$ if we assume initial fields
which are homogeneous and Gaussian with initial spectra of the form
\begin{equation}
E_{\rm tot}(k,t)=\alpha^2 |k|^ne_0(k),~~{\rm and}~~F(k,t)=\beta^2 |k|^nf_0(k).
\end{equation}
Here $n$ is the spectral index and $e_0(k)$ is an even and non-negative 
function with $e_0(0)=1$ assumed to be even and decreasing faster than any 
power of $k$ at infinity. The function $f_0(k)$ has similar properties.  
The energy spectrum at times different from zero can now be analyzed
completely as in \cite{frisch}. For example, for $1<n<2$ the spectrum 
$E_{\rm tot}(k,t)$ has three scaling regions. The first is for very small 
$k$'s, where the large eddies are conserved and the behavior agrees with the
original $|k|^n$ with a time-independent constant. The second region is a
$k^2$ region, and the third region is characterized by a behavior $k^{-2}$,
associated with the shocks. The switching from the first to the second
region occurs for a $k$ value around $t^{-1/(2(2-n))}$, whereas the shift 
to the last region occurs around $1/\sqrt{t}$, except for logarithmic 
corrections. For $-1<n<1$ there is no inner region and the spectrum develops 
in a self-similar fashion. For a much more complete description, we refer 
to the original paper \cite{frisch}. It would be of interest to see if
somewhat similar results are valid in higher dimensional MHD. It should
also be emphasized that eqs. (\ref{v+B}) and (\ref{v-B}) are the
independent equations, and hence one has the possibilty to study more
general situations than those discussed in the previous literature. For
example, the fields $v_y+B_x$ and $v_y-B_x$ may be started out with different 
random initial fields and their spectra will then develop in different ways.
We have already seen an example of this phenomenon in the perturbative 
calculation. 

Concerning the use of results obtained from the study of the Burgers
equation we mention that in eq. (\ref{v}) one can add a forcing term $f$ on the
right hand side. The very interesting study of the forced Burgers equation in 
ref. \cite{weinan} can then be used in eqs. (\ref{v+B}) and (\ref{v-B}),
which would have $f$ on the right hand side. The master equation
for the probability density functions
of $v_y+B_x$ and $v_y-B_x$, their differences and gradients, can then be 
derived as in \cite{weinan}. Again we refer to the literature \cite{weinan}
for more information.

We now return to the question as to whether our proposed equations
have some physical relevance, disregarding the obvious restrictions due
to the low dimensional structure. Our approach has the property that 
although the fields depend only on one dimension $y$, the magnetic field 
$B_x(y)$ points in a different direction. Thus we can construct an initial 
state where the magnetic field in the $x-$direction is localized in $y$
and ask how this field propagate from the equations of motion. We take the 
initial field to be
\begin{eqnarray}
&&B_x(y,0)=B_0={\rm const.}~~{\rm for}~~-L<y<L,\nonumber \\
&&B_x(y,0)=0~~{\rm otherwise,}\nonumber \\
&&v_y(y,0)=0~{\rm for~all~}y,
\end{eqnarray}
which is a small scale localized field if $L$ is not too large.
This field is a rudimentary version of a flux tube, which realistically
would be a magnetic field locally pointing e.g. in the $x-$direction with
a cross section in the $y,z-$plane. In our case this cross section 
degenerates to a line segment $-L<y<L$. Ignoring diffusion, the time
dependent solution can be obtained from eqs. (\ref{1}) and (\ref{2}),
\begin{eqnarray}
&&B_x(y,t)=\frac{1}{2}B_0,~v_y(y,t)=\frac{1}{2}B_0\nonumber \\
&&{\rm for}~~-L+tB_0<y<L+tB_0,\nonumber \\
&&B_x(y,t)=\frac{1}{2}B_0,~v_y(y,t)=-\frac{1}{2}B_0\nonumber \\
&&{\rm for}~~-L-tB_0<y<L-tB_0,\nonumber \\
&&B_x(y,t)=v_y(y,t)=0~~{\rm outside~these~intervals.}
\end{eqnarray}
Thus, the initial flux ``tube'' decays into two new tubes, which move
towards the left and the right. Also, half of the initial magnetic energy is
converted to kinetic energy, and equipartition is actually obtained. We see
that the original localized configuration turns into a less localized
configuration. This example can be generalized to more complicated initial
localized fields where $B_x$ has different values in different nearby
$y-$intervals. In such cases again the $B_x-$fields in each interval start
to split and move out to larger distances to the left and the right. In 
this way a fairly localized initial magnetic field will end up as a rather 
delocalized state, moving energy from small to large distances. This is
similar to the phenomenon of an inverse cascade in higher dimensions. 
This effect has been observed in two and three dimensions in realistic
simulations of MHD. Since this phenomenon occurs in our reduced model
one can say that our  model given by (\ref{v}) and
(\ref{B}) has some physical relevance. However, in three (but not two)
dimensions the inverse cascade is usually linked to helicity, which does not 
exist in our case for obvious reasons.
 
There are of course very important differences between MHD in one and in
higher dimensions. For example in the latter case (differential) rotation is 
possible and one can have the dynamo effect. Further, important topological
changes of flux tubes are possible. Also, the higher dimensional
MHD equations exhibit genuine chaotic behavior \cite{leshouches}, which
can only be simulated to some extent in 1+1 dimension by having random 
initial fields.

\vskip0.3cm

I thank Axel Brandenburg for telling me about the work in reference 
\cite{leshouches}. I also thank the referee for some interesting comments.

\end{multicols}

\end{document}